\documentstyle[aps,prl,preprint,psfig]{revtex}

\begin{document}

\draft
\title{Boson induced $s$-wave pairing in dilute boson-fermion mixtures}

\author{L. Viverit}
\address{Dipartimento di Fisica, Universit\`a di Milano, via Celoria 16,
20133 Milano, Italia}

\maketitle

\begin{abstract}
We show that in dilute boson-fermion mixtures with fermions in two 
internal states,
even when the bare fermion-fermion interaction is repulsive, the
exchange of density fluctuations of the Bose condensate may lead to 
an effective fermion-fermion attraction, and thus to a Cooper 
instability in the s-wave channel.
We give an analytical method to derive the associated $T_c$
in the limit where the phonon branch of the Bogoliubov
excitation spectrum of the bosons is important. 
We find a $T_c$ of the same order 
as for a pure Fermi gas with bare attraction. 
\end{abstract}


\newpage

The interest in the effective interaction between fermions in
boson-fermion mixtures is not new. Already in the 1960's Pines 
\cite{Pines} suggested that the effective interaction 
between $^3$He atoms in solution in superfluid $^4$He
could be attractive due to the exchange of density 
fluctuations of bosonic background. This attraction was then
observed
experimentally by Edwards et al. \cite{Edwards65} and Anderson et al.
\cite{Anderson66}.
From the experimental data Bardeen, Baym and Pines
\cite{BBP} estimated the expected
critical temperature for Cooper pairing both in the $s$-wave and
$p$-wave channel.
The situation is similar to that in ordinary superconductors 
where the effective attraction between electrons is well established
to be caused by the exchange of lattice phonons \cite{deGennes}.

The renewed interest in the issue stems from the recent availability
of trapped atomic gases at the ultralow temperatures required
for quantum phenomena to be relevant.
After Bose-Einstein condensation (BEC) was 
observed, huge steps have been made also in the cooling of fermions
\cite{DeMarco} --~\cite{Schreck1}. One way to obtain the cooling 
\cite{Schreck1}
has been to mix $^6$Li atoms (fermions) with $^7$Li atoms (bosons),
and to proceed with the standard technique of evaporative cooling on
the latter so as to cool the former `simpathetically', 
i.e. indirectly, by simple thermal contact. 
At the end of the process a stable nearly pure
BEC of $^7$Li on top of a cloud of degenerate
$^6$Li has been observed. It is therefore timely to calculate
the effective interaction between fermions in a dilute boson-fermion
mixture when fluctuations of the BEC are taken into account, 
and to study the conquences on Cooper pairing.

In two recent works Heiselberg, et al. \cite{Viverit00} and
Bijlsma, et al. \cite{Stoof00} 
found the same expression for the 
boson-induced contribution to the interaction,
but proceeded in different ways to analyze the effects on $s$-wave pairing.
The equivalence of the two approaches in dealing 
with the latter problem
is not obvious since the results were given in very different forms,
one analytical \cite{Viverit00} and the other numerical \cite{Stoof00}.
Moreover, their applicability is, for ref.~\cite{Viverit00} limited to
the case of an already attractive bare fermion-fermion interaction, and for
ref.~\cite{Stoof00} limited by the lack of transparency and physical
understanding of numerical calculations.
In this paper a third way to treat the problem is developed.
The idea is to give an approximate, but under usual circumstances accurate,
analytical solution to the numerical approach of ref. \cite{Stoof00}.
This applies also to the case of a bare fermion-fermion repulsion,
and we prove that, as conjectured in \cite{Stoof00} and as one expects, 
even if the bare interaction is repulsive the
effective one may be attractive when the boson-induced 
interaction is taken into account. In addition to that we find
an elegant and simple expression for the associated BCS critical temperature,
which contrary to the case of $^3$He-$^4$He mixtures is fully determined,
also in the prefactor.
Since our approach is valid for bare fermion-fermion attraction as well,
we also show that it gives, within its range of
applicability, the same results as were found in ref.~\cite{Viverit00},
and thus prove the consistency of the three methods.

When the gases of the mixture are dilute all interactions
can be described by one parameter each: the appropriate scattering 
length \cite{Landau3,Landau9}. 
In a mixture of fermions in two different internal states 1 and 2 
(with the same mass), and bosons
in one internal state only, the scattering lengths 
$a_{12}$, $a_{BB}$, $a_{B1}$ and $a_{B2}$, characterize 
all the relevant interactions, since the Pauli principle
allows us to ignore interactions between identical fermions at the
temperatures and densities of interest here. In the following we 
shall suppose, 
without loss of generality, that $a_{B1}=a_{B2}=a_{BF}$, 
and rename $a_{12}=a_{FF}$.
Often in the text pseudopotentials will be used instead of scattering
lengths. They are defined as follows:
$U_{FF}= 4 \pi\hbar^2 a_{FF} / m_{F}$, $U_{BB} = 4 \pi\hbar^2
a_{BB} / m_{B}$, and $U_{BF}= 4 \pi \hbar^2a_{BF} / m_{BF} $, where
$m_{BF} = 2 m_B m_F / (m_B+m_F)$ is twice the reduced mass for
a boson with mass $m_B$ and a fermion with mass $m_F$.
Since we are interested in $s$-wave pairing we shall
suppose that the densities of 1 and 2 fermions are the same, $n_1=n_2=n_F$, 
and therefore also their Fermi energies and momenta, 
$\epsilon_F$ and $\hbar k_F$.
Moreover we shall introduce in the theory small parameters of the type
$k_F a$ and $n_B^{1/3} a$, where by $a$ we indicated a generic 
length of the order of a scattering length and by $n_B$ the boson 
density.

In a mixture the effective interaction between a spin 1 and a spin 2 fermion
is the sum of the direct one $U_{FF}$, the one arising from polarization  
of the bosonic medium $U_{FBF}$, 
and that due to polarization of the fermionic medium
itself $U_{FFF}$. We shall show that under suitable conditions
the first two interactions can be 
of the same order, and it is therefore essential to consider both
of them to predict whether the system undergoes 
Cooper pairing or not, and if so what is the critical temperature
$T_{c}$. 
The third one is instead of order $a(k_Fa)$ and is important for
the renomalization of the prefactor in the expression for the BCS critical 
temperature. The way to deal with $U_{FFF}$
has been shown in ref.~\cite{Gorkov} (see also~\cite{Viverit00}). 
There it was found that it causes a decrease in
the prefactor of $T_{c}$ by a factor $(4e)^{1/3}$, and
we take this for granted here.

The boson-induced interaction between fermions introduced in
refs. \cite{Viverit00,Stoof00} is
\begin{equation}
\label{bindint}
U_{FBF}({\bf q},\omega)=U_{BF}^2\chi({\bf q},\omega),
\end{equation}
where the boson density-density response function in the 
Bogoliubov approximation is given by
\begin{equation}
\label{ddrf}
\chi({\bf q},\omega)=\frac{n_B\hbar^2 q^2/m_B}{(\hbar\omega)^2-\epsilon_q^0
(\epsilon_q^0+2n_B U_{BB})},
\end{equation}
$\epsilon_q^0=\hbar^2q^2/2m_B$ and $q=|{\bf q}|$.
Since we are considering dilute Bose and Fermi gases, we
have neglected the renormalization of 
$\chi(q,\omega)$ due to the presence of the fermions, this is correct to the
lowest order in the gas parameter.
We used the zero temperature response function
because the BCS critical temperature is much smaller than the boson 
condensation one if $n_B\gtrsim n_F$.

The interaction in Eq. (\ref{bindint}) provides an 
attraction between two particles at the
Fermi level, since in that case $\omega=0$ and $U_{FBF}(q,0)<0$.
We remind the reader that in uniform systems $U_{BB}>0$ 
is required for stability of the mixture \cite{Viverit99}.

At this point one has to analyze two possibilities.
The bare fermion-fermion interaction can either be attractive
$U_{FF}<0$ or repulsive $U_{FF}>0$. In the former case 
the gas would undergo pairing even in the absence of bosons at 
the critical temperature $T_{c,0}$.
In the presence of bosons the direct and induced 
contributions add up constructively to a
stronger attractive interaction, and the BCS critical temperature rises.
This possibility was studied in detail in ref. \cite{Viverit00}. 
If $U_{FF}>0$ instead, the Fermi gas in not unstable to pairing
without the bosons. When the bosons are added however, 
if the attractive boson-induced interaction at the Fermi
surface is stronger than the bare repulsion, 
the total {\em effective} interaction 
$U_{FF}+U_{FBF}(q,0)$ is attractive, and the 
gas becomes unstable to pairing. 
One can immediately recognize a mechanism at work
completely analogous to that of electrons in superconductors.
Just as in superconductors the induced interaction depends on the energy
exchanged, and is attractive only in a band centered about the Fermi
surface \cite{deGennes}.
Obtaining a solution for $T_c$ for arbitrary densities is complicated. 
As is well known however \cite{BBP,Stoof00}
if $v_F\ll s$, where $s=(n_BU_{BB}/m_B)^{1/2}$ is the sound velocity in the
Bose gas, retardation effects can be 
neglected and  $\omega$ can be set to zero always. The induced interaction is 
thus attractive in the whole Fermi sphere, and we seek a solution
to the problem under this assumption.

The boson induced interaction when $v_F\ll s$ is then
\begin{equation}
\label{eq:bosindpot}
U_{FBF}(q)=-\frac{U_{BF}^2}{U_{BB}}\cdot 
\frac{1}{1+(\hbar q/2m_B s)^2}.
\end{equation}
Notice that if $m_B\gtrsim m_F$, since the typical momentum
exchanged in an interaction is  $\hbar q\sim m_F v_F \ll 
2m_B s$,
we expect $U_{FBF}(q)\simeq -U_{BF}^2/U_{BB}$, i.e. a
constant independent of $q$.

We now want to properly take into account both bare and induced interactions 
to determine $T_c$.
According to Emery \cite{Emery} if the fermions with opposite spins 
interact via a potential $U(r)$, and if $\tan \delta_0(k_F)>0$, where
$\delta_0(k_F)$ is the associated $l=0$ phase shift
evaluated at the Fermi wavenumber, the system undergoes $s$-wave pairing 
at the critical temperature
\begin{equation}
  \label{eq:tbcsemery}
k_BT_{c}=\frac{\gamma}{\pi}\left(\frac{2}{e}\right)^{7/3}
\epsilon_F\; e^{-\pi/2\tan\delta_0(k_F)}.
\end{equation}
In writing Eq. (\ref{eq:tbcsemery}) we have already included the correction
to the prefactor due to the polarization of the fermions.

In the case of a pure two-species Fermi gas with bare attraction,
$U(r)$ is the bare potential 
(with associated scattering length and pseudopotential
$a_{FF}<0$ and $U_{FF}$ respectively), and since by assumption 
$k_F|a_{FF}|\ll 1$, then
$\tan\delta_0(k_F)\simeq -k_Fa_{FF}$, and Eq. (\ref{eq:tbcsemery}) reduces 
to the well know formula \cite{Gorkov} 
 \begin{equation}
  \label{eq:tbcsgorkov}
k_BT_{c}=\frac{\gamma}{\pi}\left(\frac{2}{e}\right)^{7/3}
\epsilon_F\; e^{\pi/2k_Fa_{FF}}.
\end{equation}

In a mixture on the other hand $U(r)$ changes.
This approach for finding the effects of the boson-induced interaction
has been used by Stoof and co-workers \cite{Stoof00}.
To apply Eq. (\ref{eq:tbcsemery}) to our case
one has to take the Fourier transform into real space coordinates
of (\ref{eq:bosindpot})
\begin{equation}
  \label{eq:yukawa}
U_{FBF}(r)=-\frac{U_{BF}^2}{4\pi U_{BB}\xi_B^2}\cdot\frac{1}{r}\,
e^{-r/\xi_B},
\end{equation}
where $\xi_B=\hbar/2m_Bs$ is the boson coherence length.
The $1/r$ divergence at $r=0$ is artificial, since the potential must
in any case be cut off at a distance $r_0$ of the order of a scattering 
length. $U_{FBF}(r)$ is a Yukawa potential with range 
$\xi_B=a_{BB}\,(16\pi\; n_B a_{BB}^3)^{-1/2}$
which is much greater than $a_{BB}$ if the gas is dilute.

The fermions interact both via $U_{\rm bare}(r)$ and $U_{FBF}(r)$ 
so that the total interaction potential is given by
$U_{\rm tot}(r)=U_{\rm bare}(r)+U_{FBF}(r)$.
The aim is then to calculate the $s$-wave phase shift due to the total 
potential $U_{\rm tot}(r)$. In principle this is a difficult problem,
since one should solve the radial Schr\"odinger equation
\begin{equation}
  \label{eq:schreq}
  \left(-\frac{d^2}{dr^2}+\frac{m_F}{\hbar^2}
U_{\rm tot}(r)-k^2\right)
u_0^{\rm tot}(r;k)=0,
\end{equation}
where, however, the bare potential is not known, as only the
scattering length $a_{FF}$ is the measured quantity.
Fortunately it is not necessary to know also the details in $r$
of $U_{\rm bare}$.
Eq. (\ref{eq:schreq}) can be largely simplified by noticing that
$U_{\rm bare}$ and $U_{FBF}$ act on two different length
scales. The first one from $r=0$ to $r\sim r_0$ and the second one from
$r\sim r_0$ to $r\sim \xi_{B}\gg r_0$.
One can then just solve Eq. (\ref{eq:schreq}) for $r>r_0$, with 
$U_{\rm tot}(r)=U_{FBF}(r)$,
and introduce a boundary condition on $u_0^{\rm tot}(r;k)$ at $r=r_0$, 
which accounts for the phase
shift due to the bare potential. This is due to the fact that by
the time the wave function reaches the region of distances where 
$U_{FBF}(r)$ is relevant, the bare potential has stopped acting 
and the wave function has recovered its sinusoidal form with phase shift
$\delta_0^{\rm bare}(k)$.
  
Using these replacements Eq. (\ref{eq:schreq}) has been solved numerically by 
Bijlsma et al. \cite{Stoof00} for various
sample elements and densities, and the results for 
$\delta_0^{\rm tot}(k)$ are in their publication. 
Whenever $\tan\delta_0^{\rm tot}(k_F)>0$ the 
critical temperature is found by replacing the value obtained 
into Eq. (\ref{eq:tbcsemery}).

We shall now show that if $k_F\xi_B\ll 1$ (a condition automatically
satisfied if $m_B\gtrsim m_F$, since by assumption $v_F\ll s$) 
$\tan\delta_0^{\rm tot}(k_F)$ can in fact be found analytically, and
the final result is remarkably simple.

It is well known, see for instance \cite{Joachain}, 
that given any two potentials
$U^{(1)}(r)$ and $U^{(2)}(r)$ for the interaction of two particles 
with reduced mass $m_{\rm red}$, and given the
solutions with wavenumber $k$ to the corresponding 
radial Schr\"odinger equations: $u_l^{(1)}(r;k)$ and $u_l^{(2)}(r;k)$,
the associated phase shifts $\delta_l^{(1)}(k)$ and
$\delta_l^{(2)}(k)$ are related by
\begin{eqnarray}
\nonumber
\tan\delta_l^{(1)}(k)&-&\tan\delta_l^{(2)}(k)=-k\frac{2m_{\rm red}}{\hbar^2}
\int_0^{\infty} u_l^{(2)}(r;k)\\
&\times&[U^{(1)}(r)-U^{(2)}(r)]\;u_l^{(1)}(r;k)\, dr.
\label{eq:wronsk71}
\end{eqnarray}
This formula can be applied to our case by letting
$U^{(1)}(r)=U_{\rm tot}(r)$ and 
$U^{(2)}(r)=U_{\rm bare}(r)$, $m_{\rm red}=m_F/2$, and $k=k_F$. 
For the $l=0$ channel then
\begin{eqnarray}
\label{eq:wronsk8}
\tan\delta_0^{\rm tot}(k_F)&=&\tan\delta_0^{\rm bare}(k_F)-k_F\frac{m_F}{\hbar^2}\\
\nonumber
&\times&\int_0^{\infty}u_0^{\rm bare}(r;k_F)U_{FBF}(r)u_0^{\rm tot}(r;k_F),
\end{eqnarray}
and since $k_F|a_{FF}|\ll 1$, the bare phase shift is
\begin{displaymath}
\tan\delta_0^{\rm bare}(k_F)\simeq-k_Fa_{FF}.
\end{displaymath}

We can now use the special form of our potentials. 
Because $U_{FBF}(r)$ is zero for  $r\lesssim r_0$,
the relevant lower limit of the integral in Eq. (\ref{eq:wronsk8}) 
is really $r_0$. But for $r>r_0$ we can replace $u_0^{\rm bare}$ 
by its asymptotic value for $r\rightarrow \infty$ since the bare 
potential has by then decayed, thus 
$u_0^{\rm bare}(r;k_F)\simeq k_F^{-1}[\sin(k_Fr)-k_Fa_{FF}\cos(k_Fr)]$.
If in addition $k_F\xi_B\ll 1$, we can approximate 
$u_0^{\rm bare}(r;k_F)\simeq r-a_{FF}$.  

Later we shall prove that if the gases are dilute
and for typical values of the parameters, the potential $U_{FBF}(r)$ is
shallow, i.e. not strong enough to form `bound states'. 
Tjinhus we can to a first approximation (Born) also let
$u_0^{\rm tot}(r;k)=u_0^{\rm bare}(r;k)$. This yields
\begin{equation}
\tan\delta_0^{\rm tot}(k_F)=-k_Fa_{FF}-A,
\end{equation}
with
\begin{eqnarray}
\label{wronsk9}
A&=&-\frac{m_Fk_F}{4\pi\hbar^2}\frac{U_{BF}^2}{U_{BB}\xi_B^2}
\int_{r_0}^{\infty}dr(r-a_{FF})^2 \frac{1}{r}\,e^{-r/\xi_B}\\
\nonumber
&=&-\frac{k_Fa_{FF}U_{BF}^2}{U_{BB}U_{FF}}
\left\{ e^{-\tilde{r}_0} (1+\tilde{r}_0) -2\tilde{a}_{FF} 
e^{-\tilde{r}_0}\right.\\
\nonumber
&+&\left.\tilde{a}_{FF}^2\int_{\tilde{r}_0}^{\infty}d x 
\frac{1}{x}\,e^{-x}\right\},
\end{eqnarray}
$\tilde{r}_0=r_0/\xi_B$, and $\tilde{a}_{FF}=a_{FF}/\xi_B$. 

The boson coherence length
is in general much larger than the cut-off $r_0$ which in turn is of order
$|a_{FF}|$. 
Therefore $\tilde{r}_0$ and $\tilde{a}_{FF}$ can be set equal to zero
in the first two terms in the curly brackets. 
The last integral is dominated by the logarithmic divergence 
and it goes like $\sim \tilde{a}_{FF}^2\ln \tilde{r}_0$,
but since $\tilde{a}_{FF}\sim\tilde{r}_0\ll 1$ also the last 
term can be set to zero. 
The final result is therefore simply
\begin{equation}
  \label{eq:wronsk03}
\tan\delta_0^{\rm tot}(k_F)=-k_Fa_{FF}\left(1-\frac{U_{BF}^2}{U_{BB}U_{FF}}
\right). 
\end{equation}

In order for the system to condense we need to have
$\tan\delta_0^{\rm tot}(k_F)>0$. This is always the case if $a_{FF}<0$,
and we recover the result of ref. \cite{Viverit00}, in the limit
$k_F\xi_B\ll 1$. 
Notice that when $k_F\xi_B\sim 1$ we cannot expand 
$u^{\rm bare}_0$ as we did anymore and we expect corrections
to our result. However, the boson-induced attraction is maximized in the
limit $k_F\xi_B\ll 1$ since it is the phonon branch of the 
Bogoliubov spectrum that provides most of the attraction, and that is why 
it is satisfactory for the time being to consider only this limit. 

Moreover, what we found extends what is reported in ref. \cite{Viverit00}, 
since we can also see that when the bare interaction is repulsive, if
$U_{BF}^2/U_{BB}U_{FF}>1$ a system which would
normally be stable undergoes a BCS transition due to the boson induced
interaction.

Our results are also consistent with those in ref. \cite{Stoof00} since
they were found using a limiting procedure of the same approach,
but the values of the phase shifts are here given analytically,
allowing a deeper physical understanding of their significance, and
showing a simplicity which was hidden by the numerics. 
To be sure we also have solved the problem numerically and found the
solutions given in ref. \cite{Stoof00}.
We have thus checked that they coincide with the present results
in the limit $k_F\xi_B\ll 1$.

In all cases in which $\lambda$ is negative 
the critical temperature is given by
\begin{equation}
  \label{eq:tbcswronsk}
k_BT_{c}=\frac{\gamma}{\pi}\left(\frac{2}{e}\right)^{7/3}
\epsilon_F\; e^{1/\lambda},
\end{equation}
where
\begin{equation}
  \label{eq:lambdawronsk2}
\lambda=N(0)U_{FF}\left\{1-\frac{U_{BF}^2}{U_{BB}U_{FF}}\right\},
\end{equation}
and $N(0)=m_Fk_F/(2\pi^2\hbar^2)$. This shows,
as we already anticipated, that the
critical temperature in the case of boson-induced pairing is of
the same order as that of pairing with bare fermion-fermion attraction,
and can be very large if $U_{BF}^2/U_{BB}U_{FF}\gg 1$. The condition
can be achieved by a suitable choice of elements. At this
time it is difficult to suggest an appropriate choice since
the scattering lengths for collisions between bosonic and fermionic 
atoms are mostly being studied at the time of writing, and precisely 
in view of the present developments to which this
work is a contribution.
We also point out that an interesting consequence of 
(\ref{eq:tbcswronsk}) and (\ref{eq:lambdawronsk2}) is that,
so long as $k_F\xi_B\ll 1$, the new critical temperature is
independent of the boson density. The strong 
density dependence in the plots
of ref. \cite{Stoof00} is explained by the fact that the regime 
$k_F\xi_B\gtrsim$ is also probed.

In the derivation above we have used the Born approximation.
This is valid only if the potential is sufficiently shallow.
But, as we see from Eq. (\ref{eq:yukawa}), the depth of the 
induced potential 
depends linearly on $n_B$ through the square of the coherence length at 
the denominator.
At low boson densities the potential is very weak and all the
considerations above are certainly valid. But when $n_B$ is large enough
for the induced potential to be able to host a `bound state' they fail,
and the new scattering length and critical temperature may depend
dramatically on $n_B$ as the calculations in ref. \cite{Stoof00}
indicate. This regime is
interesting and worth studing in greater detail, but is
beyond the scope of this work.

By a simple argument we may estimate the highest boson density allowed
for our model to apply. 
The typical potential energy of a particle confined in the potential 
(\ref{eq:yukawa}) is, apart from the sign, 
$E_p\sim U_{BF}^2/4\pi U_{BB}\xi_B^3$, and
the kinetic one $E_k\sim \hbar^2/m_F\xi_B^2$. To be safe then we
need to require $E_p/E_k\ll 1$, which implies
\begin{equation}
  \label{eq:safe1}
n_B^{1/3} a_{BB}\ll \frac{1}{(16\pi)^{1/3}}
\left(\frac{U_{BB}}{U_{BF}}\right)^{4/3}
\left(\frac{m_B}{m_F}\right)^{2/3}.
\end{equation}
Thus if the scattering lengths and the masses are approximately the same,
as in typical conditions, it is enough to require that the boson gas 
is dilute for the Born approximation to apply, but specific
checks may be necessary especially for large $U_{BF}$.
Recall that the boson density cannot be too low though if
the condition $k_F\xi_B\ll 1$ also has to be fulfilled.

In conclusion we have shown that the boson-induced interaction 
in a boson-fermion mixture can
cause $s$-wave Cooper pairing in a Fermi gas with bare repulsion. We 
have calculated the associated critical temperature in the limit
$k_F\xi_B\ll 1$ where the highly efficient attraction due to the 
phonon branch of the 
Bogoliubov spectrum is important and found that $T_c$
is of the same order as for $s$-wave pairing in a gas with bare attraction. 

Finally, I express my gratitude to C. J. Pethick 
from Nordita and H. Smith from the \O rsted laboratory in Copenhagen, 
were a large part of the work was carried out.
I would also like to thank P. F. Bortignon
and R. A. Broglia for support at the time of writing.

\end{document}